# Kubernetes as an Availability Manager for Microservice Applications

Leila Abdollahi Vayghan
Engineering and Computer Science
Concordia University
Montreal, Canada
l_abdoll@encs.concordia.ca

Mohamed Aymen Saied
Engineering and Computer Science
Concordia University
Montreal, Canada
m_saied@encs.concordia.ca

Maria Toeroe
Ericsson Inc.
Montreal, Canada
maria.toeroe@ericsson.com

Ferhat Khendek
Engineering and Computer Science
Concordia University
Montreal, Canada
ferhat.khendek@concordia.ca

*Abstract*— **The move towards the microservice based architecture is well underway. In this architectural style, small and loosely coupled modules are developed, deployed, and scaled independently to compose cloud-native applications. However, for carrier-grade service providers to migrate to the microservices architectural style, availability remains a concern. Kubernetes is an open source platform that defines a set of building blocks which collectively provide mechanisms for deploying, maintaining, scaling, and healing containerized microservices. Thus, Kubernetes hides the complexity of microservice orchestration while managing their availability. In a preliminary work we evaluated Kubernetes, using its default configuration, from the availability perspective in a private cloud settings. In this paper, we investigate more architectures and conduct more experiments to evaluate the availability that Kubernetes delivers for its managed microservices. We present different architectures for public and private clouds. We evaluate the availability achievable through the healing capability of Kubernetes. We investigate the impact of adding redundancy on the availability of microservice based applications. We conduct experiments under the default configuration of Kubernetes as well as under its most responsive one. We also perform a comparative evaluation with the Availability Management Framework (AMF), which is a proven solution as a middleware service for managing high-availability. The results of our investigations show that in certain cases, the service outage for applications managed with Kubernetes is significantly high.**

## I. INTRODUCTION

During the past decade, the computing community has witnessed a migration towards the cloud [1]. In this context, the microservices architectural style [2] has drawn a substantial attention in the industry. As opposed to the monolithic architectural style, the microservices architectural style tackles the challenges of building cloud-native applications by leveraging the benefits of the cloud [3]. Although this architectural style is poised to revolutionize the IT industry, it has so far received a limited attention from academia and research communities.

Microservices [4] are a realization of the service-oriented architectural style for developing software composed of small services that can be deployed and scaled independently by fully automated deployment machinery, with minimum centralized management [2]. Microservices are built around separate business functionalities. Each microservice runs in its own process and communicates through lightweight mechanisms, often using APIs [3]. Microservices address the drawbacks of monolithic applications. They are small and can restart faster at the time of upgrade or failure recovery. Microservices are loosely coupled, and failure of one microservice will not affect other microservices of the system. The fine granularity of this architectural style makes the scaling more flexible and more efficient as each microservice can evolve at its own pace.

To leverage all these benefits, one needs to use technologies aligned with the characteristics of this architectural style. Containerization is the technology that enables virtualization at the operating system level [5]. Containers are lightweight and portable. Therefore, they are suitable for creating microservices. Docker [6] is the leading container platform that packages code and dependencies together and ships them as a container image. Since containers are isolated, they are not aware of each other. Thus, there is a need for an orchestration platform to manage the deployment of containers. Kubernetes [7] is an open-source platform that enables the automated deployment, scaling, and management for containerized applications. Kubernetes relieves application developers from the complexity of implementing their application's resiliency. Therefore, it has become a popular platform for deploying microservice based applications.

The move towards the microservice based architectures is well underway. However, as an important quality attribute for carrier grade service for instance, the availability remains a concern. Availability is a non-functional characteristics defined as the amount of service outage over a period of time [8]. High availability is achieved when the system is available at least 99.999% of the time. Therefore, the total downtime allowed in one year for highly available systems is around 5 minutes [9]. Some characteristics of microservices and containers such as being small and lightweight would naturally contribute to improve the availability [10]. Kubernetes provides healing for its managed microservice based applications [7]. The healing capability of Kubernetes consists of restarting the failed containers and replacing or rescheduling containers when their hosts fail. The healing capability is also responsible of advertising about the unhealthy containers until they are ready again. These features would also naturally improve the availability of the services provided by the applications deployed with Kubernetes. The question is what is the availability rendered by these applications?

In this paper, we are interested in evaluating microservice based applications from the availability perspective, since our ultimate goal is to enable high availability for microservices. As a follow up to [11] for an initial setup in a private cloud and the default Kubernetes configuration for healing, we have investigated other architectures, configurations, and conducted a series of experiments with Kubernetes and measured the outage times for different failure scenarios. The goal was to answer the following research questions (RQ):

- **RQ1**: What is the level of availability that Kubernetes can support for its managed microservices solely through its healing features?
- **RQ2**: What is the impact of adding redundancy on the availability achievable with Kubernetes?
- **RQ3**: What is the availability achievable with Kubernetes under its most responsive configuration?
- **RQ4**: How does the availability achievable with Kubernetes compare to existing solutions?

We conducted our experiments under the default configuration of Kubernetes as well as its most responsive one. To better position and characterize the obtained results we opted for a comparison with an existing solution for availability management, the Availability Management Framework (AMF) [12], a proven middleware service for high availability (HA) management.

The rest of the paper is organized as follows. Section II introduces the Kubernetes' architecture components and the different architectural solutions for deploying microservice based applications with Kubernetes. In Section III we present our experiments' settings, the failure scenarios and the availability metrics. The experiments, the results and the analysis with respect to the research questions are presented in Section IV. We discuss the lessons learned and the threats to validity in Section V. In Section VI, we review the related work on microservice based applications deployed with Kubernetes from the availability perspective. We conclude in Section VII.

## II. ARCHITECTURES FOR DEPLOYING MICROSERVICE BASED APPLICATIONS WITH KUBERNETES

### A. Kubernetes Architectural Components

Kubernetes is a platform for automating the deployment and scaling of containerized applications across a cluster [7]. The Kubernetes cluster has a master-slave architecture. The nodes in a Kubernetes cluster can be either virtual or physical machines. The *master node* hosts a collection of processes to maintain the desired state of the cluster. The *slave nodes*, that we will refer to them simply as nodes, have the necessary processes to run the containers and be managed by the master [7]. The most important process running on every node is called Kubelet [7]. Kubelet runs the containers assigned to its node via Docker, periodically performs health checks on them, and reports to the master their statuses as well as the status of the node.

The smallest and simplest unit that Kubernetes deploys and manages is called a pod [7]. A pod is a collection of one or more containers and provides shared storage and network for its containers. Containers in a pod share its IP address and port space. A pod also has the specifications of how to run its containers. Customized labels can be assigned to pods to group and query them in the cluster. All this information is described in the pod template. In practice, microservices are containerized and deployed on a Kubernetes cluster as pods.

The pods in a Kubernetes cluster are deployed and managed by controllers [7]. A controller specification consists of the pod template, the desired number of replicas of that pod the controller should maintain at all times, and other information such as upgrade strategy and pods' labels. Once the controller is deployed to the cluster, it creates the desired number of pods based on the provided template and continuously maintains their number equal to the desired number. For example, when a pod fails due to a node failure, the corresponding controller will automatically create a new one on another node.

There are different types of controllers in Kubernetes and each of them is suitable for a specific purpose. For example, DaemonSet controllers run a copy of a pod on all nodes, Job controllers create a number of pods and make sure they successfully terminate, and StatefulSet controllers are used to manage stateful applications. In this paper, we focus on the deployment controller used for deploying stateless applications.

As mentioned before, a pod has its own IP address, and this IP address may change often as the pod is deleted and revived dynamically by its controller. This frequent change of IP addresses makes it impossible to keep track of the pods and communicate with them through their IP addresses. Kubernetes provides an abstraction called service [7], which defines a logical set of pods as its endpoints and a policy by which to access them. A service groups pods based on their labels and not based on their IP addresses, and so it hides the changes of IP addresses of pods from the client. A service has an intercluster virtual IP address that redirects to its endpoints either randomly or in a round robin manner.

Besides Kubelet, the other important process running on all nodes is called Kube-Proxy [7]. Kube-Proxy watches the master for information about the services created in the cluster and their endpoints. It updates the iptables of the node and adds rules to forward the requests for a service to its endpoints. When a service or its endpoint is removed, Kube-Proxy updates the iptables of the node accordingly.

Kubernetes' services can be of different types. The default type is called "Cluster IP". Services of this type are accessible only from within the cluster. The "Node Port" type of service is built on top of a Cluster IP service and exposes the service on the same port of each node of the cluster. Lastly, a "Load Balancer" type of service is exposed externally only when the cluster is running in a public cloud.

Kubernetes provides another way, called ingress, to access services from outside of the cluster [7]. An ingress is a collection of rules for inbound connections to reach certain services in the cluster that are defined as backends for the ingress. For an ingress to work, an ingress controller needs to run on the cluster. Ingress controllers are not part of Kubernetes. To have an ingress controller, one should either implement it or use one that is available, e.g. Nginx [13] or HAProxy [14].

Kubernetes hides all this complexity behind its API. Therefore, Kubernetes' users do not need to implement the required mechanisms to manage their applications' resilience. Kubernetes' users only have to interact with the API to specify the desired deployment architecture and Kubernetes will be in charge of orchestration and availability management of the application. However, users with advanced requirements such as high availability may need to dive into Kubernetes details, since the Kubernetes architectural components can be used in different ways to deploy applications in a Kubernetes cluster. For example, an application can be deployed without using services at all. In this case, a mechanism should be implemented to guarantee that each pod advertises its IP address to the rest of the pods. Moreover, Kubernetes can run in a cluster in a public or a private cloud. The architecture and the efforts needed to deploy an application in each of these platforms is different. In this paper, we discuss architectures for application deployment in both public and private clouds. These architectures are based on our understanding of Kubernetes architectural components described in [7].

### B. Deploying Containerized Applications in a Kubernetes Cluster Running in a Public Cloud

In this section, we consider a Kubernetes cluster composed of VMs running in a public cloud. Kubernetes runs on all VMs and creates a unified view of the cluster. One of the VMs is selected as the master and it is in charge of managing the nodes. As we are concerned with High Availability, we should consider an HA cluster composed of more than one master. However, such setting is still experimental and non-mature for Kubernetes. Thus, we decided to go with only one master and keep failure from the master side out of the scope of this paper. For simplicity, the application here is composed of only one microservice. The pod template for the containerized microservice as well as its desired number of replicas are included in a deployment controller specification which is deployed to the cluster. We will discuss two ways to expose services in Kubernetes clusters running in a public cloud.

*1) Service of Type Load Balancer:* An architecture for deploying applications in a Kubernetes cluster using a service of type Load Balancer in a public cloud is shown in Fig. 1. In addition to a cluster IP, services of type Load Balancer have an external IP address that is automatically set to the cloud provider's load balancer IP address. Using this external IP address, which is public, it is possible to access the pods from outside of the cluster.

*2) Ingress:* There could be more than one service that need to be exposed externally and with the previous method, one load balancer is needed for each service. On the other hand, Kubernetes' ingress resource can have multiple services as backends and minimize the number of load balancers *[7]*. In a Kubernetes cluster running in a public cloud, an ingress controller is deployed and exposed by a service of type Load

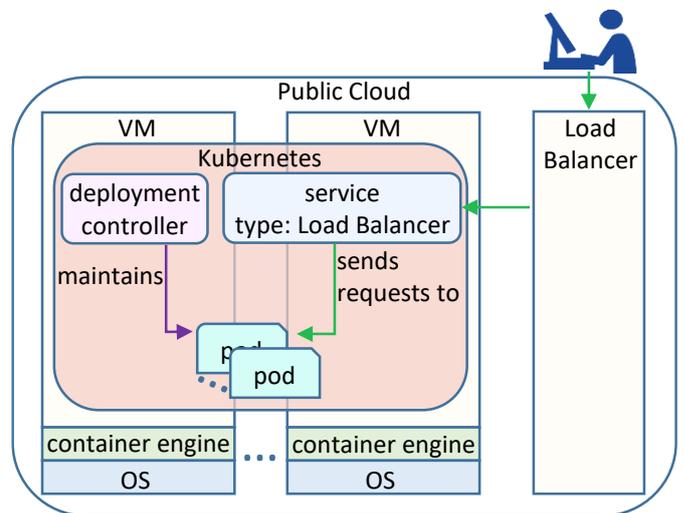

*Fig. 1. Architecture for deploying applications to Kubernetes clusters running in a public cloud.*

Balancer *[7]*. Therefore, requests for all services that are sent to the cloud provider's load balancer are received by the ingress controller and redirected to the appropriate service based on the rules defined in the ingress resource.

### C. Deploying Containerized Applications in a Kubernetes Cluster Running in a Private Cloud

As it was previously mentioned, Kubernetes is designed to run on different types of platforms. However, it is important to understand that deploying applications to a Kubernetes cluster running in a private cloud requires more effort than it does for a public cloud. The main difference is in the way of exposing the application externally. Below, we will discuss two ways of exposing applications deployed in a Kubernetes cluster running in a private cloud.

*1) External Load Balancer:* Fig. 2 depicts the architecture for exposing the service using an external load balancer. Services of type Node Port expose the service on the same port on every node in the cluster. Since it is not a good practice to expect the users to connect to the nodes directly, an external load balancer is used, which distributes the requests between the nodes and delivers them to the port on which the Node Port service is exposed. The downside of this architecture is that for each service in the cluster that needs to be exposed externally, we will need one external load balancer.

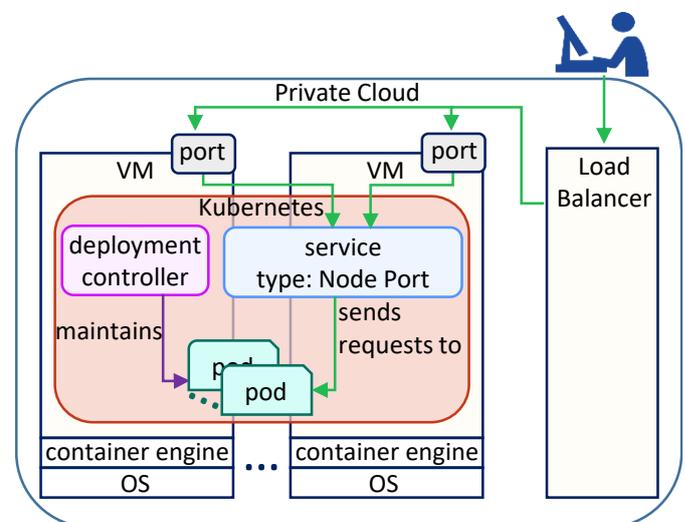

*Fig. 2. Private cloud - exposing services via external load balancers.*

*2) Ingress:* Using Kubernetes' ingress resource is a more structured way to expose services. In this case, a service of type Cluster IP is created to redirect requests to the pods and will be used as the backend of the ingress resource. Also, an ingress controller is needed in the cluster in order to redirect the incoming requests to the ingress resource, which later will be redirected to the appropriate backend service. The ingress controller is deployed as one pod using a deployment controller. We create a service of type Node Port to expose the ingress controller pod to outside of the cluster. Since it is possible to define multiple services as backends for the ingress resource, this method is more practical than the previous case in which we needed to use a load balancer for each individual service that needs to be exposed. Adapting the ingress controller to a Kubernetes cluster running in a private cloud is not an easy task and there is no sufficient documentation on how to use ingress controllers in these types of clusters. Fig. 3 shows a generic architecture with ingress exposing the service in a cluster running in a private cloud to the outside world.

Although the role of each Kubernetes architectural component is described in [15], understanding how to put them together is not intuitive. It is a time-consuming effort and requires lots of trials and errors to figure out the ways these components work together in practice. The aforementioned architectures are a result of our understanding of these components' roles and several weeks of trials.

Even though Kubernetes can run in both private and public clouds, one may notice that it is better tailored for public clouds than it is for private ones. For a Kubernetes cluster running in a public cloud, the application is automatically exposed to the outside world through a service of type Load Balancer because it can use the cloud provider's load balancer. Also, using the ingress resource for redirecting requests to multiple services is less challenging. On the other hand, in a private cloud, exposing the application to the outside world is challenging and requires more efforts. One needs to either tackle the complexity of adapting an ingress controller to expose services or use an external load balancer for each service that needs to be exposed which will not be practical for large and complex microservice based applications composed of many microservices.

### III. SETTINGS, FAILURE SCENARIOS AND METRICS

In this section, we describe the settings for the experiments, the failure scenarios we considered as well as the availability metrics. We set a Kubernetes cluster in a private cloud (Fig. 3). This cluster is composed of three VMs running on OpenStack cloud. Ubuntu 16.04 is the OS running on all VMs. Kubernetes 1.8.2 runs on all VMs and the container engine is Docker 17.09. Network Time Protocol (NTP) [16] is used for time synchronization between the nodes. The application deployed is VideoLan Client (VLC) [17]. There is one container image in the pod template, on which VLC is installed. Once a pod is deployed, an application container will be created based on this image and will start streaming from a file.

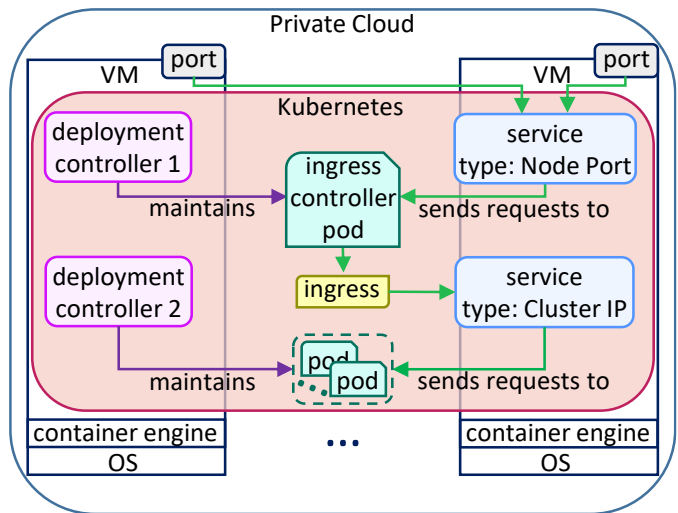

*Fig. 3. Private cloud - exposing services via ingress.*

Kubernetes offers three levels of health check and repair action for managing the availability of the deployed microservices. First, at the application level, Kubernetes ensures that the software components executing inside a container are healthy either through process health check or predefined probes. In both cases, if the Kubelet discovers a failure, the container is restarted. Second, at the pod level, Kubernetes monitors the pod failures and reacts according to the defined restart policy. Finally, at the node level, Kubernetes monitors the nodes of the cluster through its distributed daemons for node failure detection. If the node hosting a pod fails, the pod is rescheduled into another healthy node. With respect to these levels of health check, we defined three sets of failure scenarios. In the first set, the application failure is caused by the VLC container process failure. In the second set, it is due to pod container process failure, and in the third set it is caused by the node failure. For each set, we experimented with different redundancy models [18] and with both default and most responsive configuration of Kubernetes. Each scenario has been repeated 10 times and the average of the measurements are shown in Table I through Table V. All the measurements reported in this paper are in seconds.

The metrics we use to evaluate Kubernetes from availability perspective are defined below. In Fig. 4 we summarize the relations between these metrics.

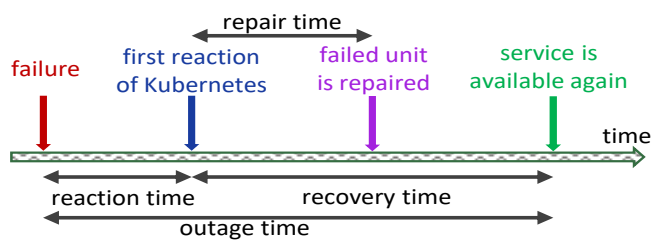

Fig. 4. Availability metrics.

TABLE I. EXPERIMENTS WITH KUBERNETES - NO-REDUNDANCY REDUNDANCY MODEL AND DEFAULT CONFIGURATION

| Failure Trigger (unit: seconds) | Reaction time | Repair time | Recovery time | Outage time |
|---|---|---|---|---|
| VLC Container Failure | 0.716 | 0.472 | 1.050 | 1.766 |
| Pod Container Failure | 0.496 | 32.570 | 31.523 | 32.019 |
| Node Failure | 38.187 | 262.542 | 262.665 | 300.852 |

**Reaction Time:** The time between the failure event we introduce and the first reaction of Kubernetes that reflects the failure event was detected.

**Repair Time:** The time between the first reaction of Kubernetes and the repair of the failed pod.

**Recovery Time:** The time between the first reaction of Kubernetes and when the service is available again.

**Outage Time:** The duration in which the service was not available. It represents the sum of the reaction time and the recovery time as shown in Fig. 4.

## IV. EXPERIMENTS, RESULTS AND ANALYSIS

In this section, we present the architectures, the experiments, the results and the analysis for answering the research question we posed in the introduction.

### A. Evaluating the Repair Actions with Default Configuration of Kubernetes for Supporting Availability (RQ1)

Fig. 5 shows the architecture for these experiments. The redundancy model in this case is No-Redundancy [18] and therefore, the number of pods in the deployment controller specification is only one.

*1) Experiments*

We evaluate the availability metrics for each of the failure scenarios under the default configuration of Kubernetes.

**Service Outage due to VLC Container Process Failure:** In this scenario, the failure is simulated by killing the VLC container process from the OS. When the VLC container crashes, the Kubelet detects the crash and brings the pod to a state where it will not receive new requests. At this time, that is the reaction time, the pod is removed from the endpoints list. Later, the Kubelet restarts the VLC container and the video will start from the beginning of the file. This time marks the repair time. Recovery time is when the pod is in the endpoints list again and is ready to receive requests.

**Service Outage due to Pod Container Process Failure:** When a pod is deployed, along with the application containers specified in its template, one extra container is created which is the pod container. Since the pod container is a process in the OS, it is possible that it crashes. In this scenario, the failure is simulated by killing the pod container process from the OS. When the pod container process is killed, the Kubelet detects that the pod container is no longer present and this marks the reaction time. When the new pod is created and its VLC container is started, the video will start streaming from the beginning of the file and we consider the pod as repaired. After, the Kubelet will add the new pod to the endpoints list and it will be ready to receive new requests, this marks the recovery time.

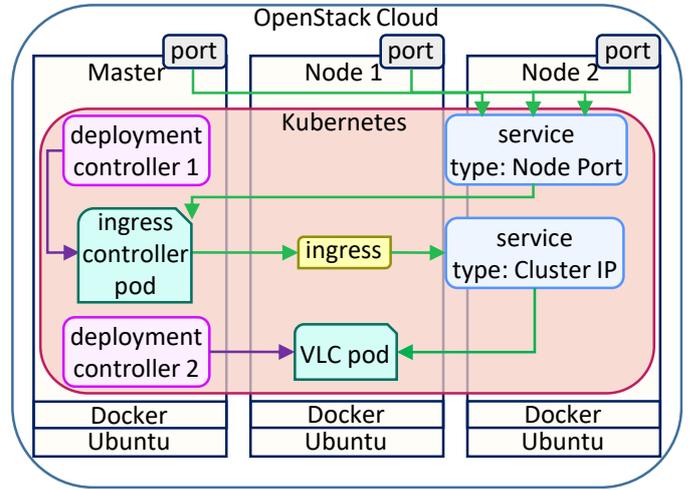

Fig. 5. Concrete architecture for deploying applications with Kubernetes - No-Redundancy redundancy model.

**Service Outage due to Node Failure:** In this scenario, a node failure is simulated by the Linux's reboot command on a VM hosting the pod. As mentioned before, the Kubelet is responsible to report the status of the node to the master, and it is the node controller of the master who detects the failure of the node. When a node hosting a pod fails, it stops sending status updates to the master and the master will mark the node as not ready after the fourth missed status update. This time is the reaction time. When the node is marked as not ready, the VLC pod on the node is scheduled for termination and after it is completely terminated a new one will be created. The repair time is when the new VLC pod is started and streaming the video. Recovery time is when the pod is added to the endpoints list of the service.

*2) Results and Analysis*

The measurements and events of this set of experiments are shown in Table I and Fig. 6, respectively. In Fig. 6, the failure of the VLC container, the pod container or the node hosting Pod1 is shown as the first event. Before this event, the IP address of Pod1 was in the endpoints list and the service was available. After the failure, the service becomes unavailable. However, since Kubernetes has not detected the failure yet, the IP address of Pod1 stays in the endpoints list. It is removed from the endpoints list at the reaction time.

In this architecture, removing the IP of Pod1 from the endpoints list as a reaction of Kubelet to the VLC container failure takes 0.716 seconds, and in the case of pod container failure it takes 0.496 seconds. However, in the case of node failure, removing the IP of Pod1 from the endpoints list, as a reaction from the node controller of the master was measured as 38.187 seconds. The reason is that with the default configuration of Kubernetes, the master takes at least 30 seconds to detect a node failure. Because the Kubelet updates the node status every 10 seconds and the master allows for four missed status updates before marking the node as not ready.

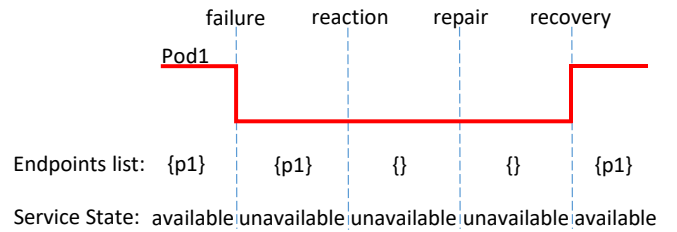

Fig. 6. Analysis of experiments with Kubernetes under the default configuration and No-Redundancy redundancy model – evaluating the repair actions.

TABLE II.      EXPERIMENTS WITH KUBERNETES – N-WAY ACTIVE
REDUNDANCY MODEL AND DEFAULT CONFIGURATION

| Failure Trigger (unit: seconds) | Reaction time | Repair time | Recovery time | Outage time |
|---|---|---|---|---|
| VLC Container Failure | 0.579 | 0.499 | 0 | 0.579 |
| Pod Container Failure | 0.696 | 30.986 | 0.034 | 0.730 |
| Node Failure | 38.554 | 262.178 | 0.028 | 38.582 |

The repair time for all scenarios is when a new pod is created again and streaming the video starts again. As observed in Table I, the repair time of the VLC container or the pod container failure scenarios differ significantly (0.472 seconds for the former and 32.570 seconds for the latter). The reason is that in the case of pod container failure, a graceful termination signal is sent to the VLC container and Docker waits 30 seconds for it to terminate. The repair process will not start unless the VLC container is terminated. For the node failure scenario, as shown in Table I, the repair time is considerably high. The reason is that with the default configuration of Kubernetes, in the case of node failure, the master waits around 260 seconds to start a new pod and recover the service. Because of these high repair times, the service outages for the pod container and node failure scenarios are significantly high, 32.019 seconds and 300.52 seconds, respectively.

### B. Evaluating the Impact of Redundancy on the Availability Support by Kubernetes (RQ2)

To investigate the impact redundancy may have on the availability support by Kubernetes, we consider the architecture in Fig. 7 where the number of pod replicas that the deployment controller maintains is increased to two. In this architecture, we have a N-Way Active redundancy model [18].

*1) Experiements*

We evaluate the availability metrics for each of the failure scenarios under the default configuration of Kubernetes with a N-Way Active redundancy model. We compare the results to the previous experiments (Section IV.A).

**Service Outage due to the VLC Container Process Failure:** In this scenario, similar to the No-Redundancy redundancy model, the reaction time is when the Kubelet detects the VLC container has crashed and removes the pod from the endpoints list. By just removing the unhealthy Pod1 from endpoints list, the service is recovered. This is because another healthy pod is still on the endpoints list and ready to serve the requests. Therefore, the reaction time for this scenario is the same as the recovery time. The repair time is when the Kubelet has restarted the crashed VLC container and the video has started streaming again.

**Service Outage due to Pod Container Process Failure:** In this scenario, same as for the No-Redundancy redundancy model architecture, the reaction time is when the Kubelet detects that the pod is no longer there. Similarly to the previous scenarios, the recovery time is when the unhealthy pod is removed from the endpoints list. The repair time is when a new pod is created and its VLC container is started and streaming the video.

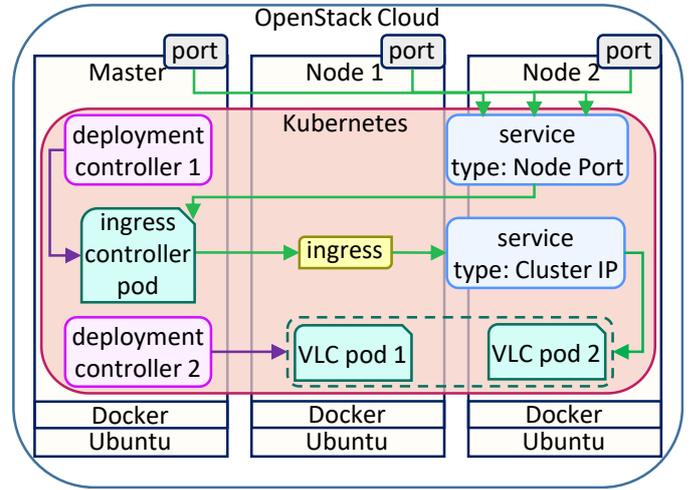

*Fig. 7. Concrete architecture for deploying applications with Kubernetes – N-Way Active redundancy model.*

**Service Outage due to Node Failure:** The reaction time in this scenario is the same as for the No-Redundancy redundancy model architecture, i.e. the time the master marks the node as not ready and schedules the pod for termination. The recovery time is when the IP of Pod1 is removed from the endpoints list. The repair time is when Pod1 is terminated and another one is created.

*2) Results and Analysis*

The measurements and the events for this set of experiments are shown in Table II and Fig. 8, respectively. The failure of the VLC container, the pod container or the node hosting Pod1 is shown in Fig. 8. Before this event, the IP addresses of Pod1 and Pod2 were in the endpoints list and the service was available. After the failure, the service is degraded. The reason is that Kubernetes has not detected the failure yet and the IP address of the failed Pod1 is still in the endpoints list. At this point, since Kubernetes still assumes that Pod1 is healthy, some requests may be redirected to Pod1. Therefore, we consider the service as degraded.

For the three failure scenarios, the measured recovery time is the same and this is when the IP of the failed Pod1 is removed from the endpoints list. However, in the case of the VLC container failure, this event marks both the first reaction to the failure as well as the recovery of the service. Therefore, for this scenario, recovery time is zero. Repair time happens later when the failed pod is completely terminated and Pod1 is created again and streaming the video.

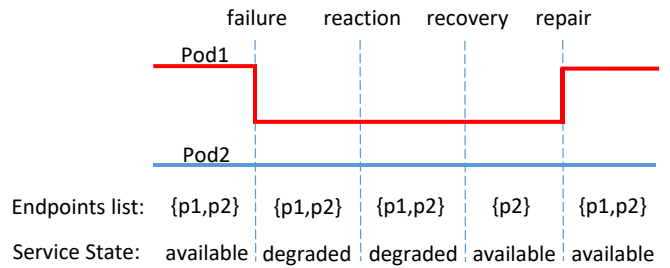

*Fig. 8. Analysis of experiments with Kubernetes under the default configuration and N-Way Active Redundancy model – evaluating the impact of redundancy.*

As shown in Table II, the measured outage time in the experiments with a N-Way Active redundancy model is significantly lower than for the No-Redundancy redundancy model. For instance, the outage times for the pod container failure and the node failure scenarios were reduced from 32.019 and 300.852 seconds to 0.730 and 38.582 seconds, respectively. The reason is that with the N-Way Active redundancy model, the recovery does not depend on the repair of the faulty unit and the service is recovered as soon as Kubernetes detects the failure. The results show that the repair actions and the healing capability of Kubernetes are not sufficient for supporting availability and adding redundancy can significantly decrease the downtime.

### C. Evaluating the Repair Actions with Most Responsive Configuration of Kubernetes for Supporting Availability (RQ3)

As observed in subsections A and B, the default configuration of Kubernetes has a significant impact on the service outage. Our analysis for the different failure scenarios has led to the identification of the aspects that need to be modified to reduce the observed outage. One aspect affecting the service outage is the graceful termination signal sent to the application container in the scenario of pod container failure. For the node failure scenarios, the frequency of node status posting by Kubelet to the master and the number of allowed missed status updates before marking a node as unhealthy are the main aspects that affect the service outage.

#### 1) Experiments

We perform two sets of experiments where Kubernetes has the most responsive configuration. In the first set, for the pod container failure, the configuration parameter for the graceful termination of pods is set to zero second. In the second set, for the node failure, the configuration parameters related to handling node failure are set to the lowest value possible (one second). We are aware of the network overhead and potential false node failure reports for the most responsive configuration. However, our goal in this experiment is to measure the best achievable availability when deploying applications with Kubernetes. These experiments were conducted with both No-Redundancy and N-Way Active redundancy model architectures (Fig. 5 and Fig. 7).

**Reconfiguring the Graceful Termination Period of Pods:** As it was mentioned, when a pod container process fails, a graceful termination signal is sent to Docker to terminate the application container which delays the repair of the pod for 30 seconds. In the No-Redundancy redundancy model, this grace period affects the recovery time, because a new pod will not be created unless the failed one completely terminates. To reduce this grace period, we updated the pod template and set the grace period to zero. We repeated the experiments for the pod container failure scenario and evaluated the impact of this change on service outage.

**Reconfiguring Node Failure Handling Parameters:** To have the most responsive Kubernetes configuration, we reconfigured the Kubelet of each node to post the node's status every second to the master. The node controller of the master was also reconfigured to read the updated statuses every second and only allow one missed status update for each node. We repeated the experiments for the node failure scenario in order to evaluate the impact of this reconfiguration on service outage.

#### 2) Results and Analysis

The results of these experiments are presented in Table III and Table IV. As it was expected, Table III shows a significant decrease in repair time which affects the service outage of experiments done with No-Redundancy redundancy model. The service outage of experiments with the N-Way Active redundancy model has not changed, as the repair time does not play a role in the service outage in this case. We observed that with the new configuration, when the pod container crashes, the time Docker gives to the application container before forcefully killing it is reduced to 2 seconds. Moreover, in Table IV, we observe significant changes in all measured metrics. With the new configuration, the master allows only one missed status update and since each node updates its status every second, the reaction time is reduced to almost one second. The repair time also is decreased from 260 seconds to around 2.5 seconds, as the master will wait only one second before starting a new pod on a healthy node. The change in the reaction time affects the service outage in all the experiments. However, the repair time affects the outage time in the case of the No-Redundancy redundancy model only.

### D. Comparing Kubernetes with Existing Solutions for Availability Management (RQ4)

To better position the availability results obtained with Kubernetes, we look into RQ4 (*How does the availability achievable with Kubernetes compare to existing solutions?*). AMF [12] is a standard middleware service for managing the availability of components based applications. It has been implemented, with other middleware services, in the OpenSAF middleware [19], a proven solution for availability management. In [20] we conducted a set of experiments for different failure scenarios with the same application, VLC. We considered the following failure scenarios, VLC process failure, VM failure and physical host failure, corresponding to VLC container failure, pod container failure and node failure, respectively. In the experiments with OpenSAF we used a No-redundancy redundancy model with two VLC applications, one active and the other one as a spare to be instantiated and take over in case of failure of the active.

TABLE III. EXPERIMENTS WITH KUBERNETES WITH CHANGED CONFIGURATION - SERVICE OUTAGE DUE TO POD CONTAINER FAILURE

| Redundancy Model (unit: seconds) | Reaction time | Repair time | Recovery time | Outage time |
|---|---|---|---|---|
| No-Redundancy | 0.708 | 3.039 | 3.337 | 4.045 |
| N-Way Active | 0.521 | 3.008 | 0.032 | 0.554 |

TABLE IV. EXPERIMENTS WITH KUBERNETES WITH CHANGED CONFIGURATION - SERVICE OUTAGE DUE TO NODE FAILURE

| Redundancy Model (unit: seconds) | Reaction time | Repair time | Recovery time | Outage time |
|---|---|---|---|---|
| No-Redundancy | 0.976 | 2.791 | 2.998 | 3.974 |
| N-Way Active | 0.849 | 2.173 | 0.022 | 0.872 |

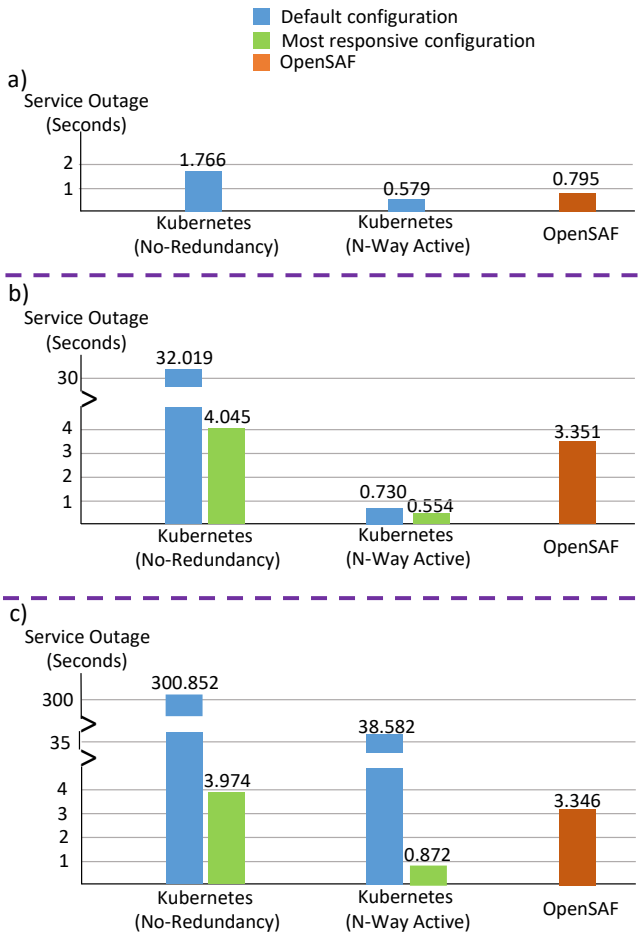

Fig. 9. Comparing Kubernetes and OpenSAF from availability perspective. a) VLC container failure scenario, b) Pod container failure scenario, c) Node failure scenario.

The results of the experiments with OpenSAF and the comparison with Kubernetes are shown in Table V and Fig. 9, respectively. We observe that in the cases of No-Redundancy redundancy model, the OpenSAF solution shows a lower outage time. Moreover, although the N-Way Active redundancy model should render a higher level of availability compared to the No-Redundancy redundancy [18], the outage time for the node failure scenario of Kubernetes with N-Way Active is still significantly higher than for OpenSAF with the No-Redundancy redundancy model. The reason for this is the default configuration of Kubernetes that leads to a late reaction time.

TABLE V. EXPERIMENTS WITH OPENSAF

| Failure Trigger (unit: seconds) | Reaction time | Repair time | Recovery time | Outage time |
|---|---|---|---|---|
| VLC Process Failure | 0.650 | - | 0.145 | 0.795 |
| VM Failure | 3.233 | - | 0.123 | 3.351 |
| Physical Host Failure | 3.229 | - | 0.118 | 3.346 |

However, with the changed configuration of Kubernetes, the outage times in Kubernetes experiments with No-Redundancy architecture are comparable to those of OpenSAF.

## V. LESSONS LEARNED AND THREATS TO VALIDITY

### A. Lessons Learned

Kubernetes supports automatic deployment and scaling of microservice based applications. Although Kubernetes can run on different cloud environments, one has to admit that its deployment in a private cloud is not as straightforward as in public clouds. Kubernetes provides availability through its repair actions. However, these are not sufficient for supporting highly availability. For example, for the node failure scenario, the outage time is about 5 minutes, which is equivalent to the amount of downtime allowed in a one-year period for a highly available system. Even after adding redundancy, we observed that the default configuration of Kubernetes still resulted in a significant service outage in the case of node failure. Although the default configuration can be changed, Kubernetes is most commonly used under its default configuration and figuring out how to reconfigure Kubernetes' reaction to node failure while avoiding network overhead and false positive reports can be complicated and requires a great effort.

### B. Threats to Validity

*1) Internal Validity:* the following internal threats can affect the validity of our results. First, all experiments were conducted in a small cluster consisting of only a master and two worker nodes. Kubernetes may behave differently in larger clusters which may impact the availability measurements presented in our experiments. Second, the availability measurements may also vary depending on the application's complexity and the collocated applications managed by Kubernetes. In our experiments, we considered a simple case of only one microservice. We understand that these factors may impact the results of our study. However, we believe that these factors can only decrease the availability of the application. The mapping of the metrics to the concrete events is the biggest threat and requires more investigation as one can map them differently, in which case all the measurements could be different. However, we believe that even with a different mapping what would change is the split between reaction and repair times and reaction and recovery times, thus, resulting still in the same outage time. We may observe a decrease in the reaction time which adds to the recovery time, or inversely but the total outage time would be the same since it represents the duration in which the service was not available.

*2) External Validity:* We only considered the case of a video streaming application. Before generalizing the results, one has to consider other types of applications even though the conducted experiments and the analysis give some indications about the availability of the applications deployed with Kurbernetes.

*3) Construct Validity:* Regarding the extent to which the observed phenomena correspond to what is intended to be observed, we see another threat related to the tools and mechanisms used in our experiments. Indeed, we rely on the timestamps reported in Kubernetes and Docker logs. However, we used NTP to synchronize the time between the nodes. Other strategies can be used and may be more meticulous, for instance, container instrumentation. However, we believe that the considered logs are precise enough for our needs, and to mitigate this threat, we cross-checked the timestamp data in the different logs (i.e. Kubernetes, Docker, and systemd journal). Moreover, to alleviate any particular result expectancies we used the same process for the different failure scenarios.

## VI. RELATED WORK

The architectural style of microservices has emerged primarily from the industry [2]. It is being adopted and investigated from different perspectives by practitioners and to a smaller extent by researchers in academia as well. In this section, we review related work focusing to the availability of microservice based architecture.

Dragoni et al. in their work [3] propose the definition of a microservice as a small and independent process that interacts by messaging. They define the microservice based architecture as a distributed application composed of microservices and discuss the impact of microservices on the quality attributes of the application. Along with performance and maintainability, they specifically discuss availability as a quality attribute which is impacted by the microservice based architecture. Emam et al. in [21] found that as the size of a service increases, it becomes more fault-prone. Since microservices are small in size, in theory, they are less fault-prone. However, Dragoni et al. argue that at integration, the system will become more fault-prone because of the complexity of launching an increasing number of microservices.

Khazaei et al. in [22] propose a microservice platform for the cloud by using a Docker technology that provisions containers based on the requests of microservice users. One of the key differences between this platform and Kubernetes is that this platform has the ability to ask for more VMs from the infrastructure when needed while Kubernetes does not. Kang et al. in [23] propose a microservice based architecture and use containers to operate and manage the cloud infrastructure services. In their architecture, each container is monitored by a sidekick container and in case of failure, recovery actions are taken. They performed some experiments and concluded that recovering from container failure is faster than recovering from VM failure.

Netto et al. in [24] believe that Kubernetes improves the availability of stateless applications but faces problems when it comes to stateful applications. They integrated a coordination service with Kubernetes to offer automatic state replication. In their architecture, all replicas of a pod execute the incoming requests while only one, which has received the request from the client, will respond. This way, the state is replicated. They evaluated some metrics such as latency and they concluded that the latency increases with the number of pod replicas.

In our study, we performed a quantitative evaluation and analysis of availability for microservice based applications deployed with Kubernetes. We considered several failure scenarios, configurations and redundancy models. We compared the results with Kubernetes to similar settings with OpenSAF to position Kubernetes from the availability perspective.

## VII. CONCLUSION AND FUTURE WORK

In this paper, we presented and compared architectures for deploying microservice based applications in Kubernetes clusters hosted in public and private clouds. Through our investigations, we learned that although it is not stated in Kubernetes' documentation [15], Kubernetes is more tailored for public clouds than for private clouds. We conducted experiments in a private cloud environment, considering different failure scenarios, configurations, redundancy models, to evaluate Kubernetes from the perspective of the availability of the applications running on Kubernetes. We analyzed the results of our experiments and found that the repair actions of Kubernetes are not sufficient for providing availability, especially high availability. For instance, the default configuration of Kubernetes results in a significant outage in the case of node failure. Kubernetes can be reconfigured to avoid this significant outage and under its most responsive configuration, outage times in Kubernetes experiments are comparable to those of OpenSAF, a proven solution for availability management. We also observed that adding redundancy can significantly decrease the downtime since the service is recovered as soon as Kubernetes detects the failure and it does not depend on the repair of the faulty unit. In our future works, we will investigate architectures for deploying stateful microservice based applications with Kubernetes. More investigations are also required to find out the impact of reconfiguring Kubernetes on network overhead and false positive node failure reports. We will also consider, the case of multiple masters and failure from the master side, as well as Kubernetes behavior in case of network partitioning.

ACKNOWLEDGMENT

This work has been partially supported by Natural Sciences and Engineering Research Council of Canada (NSERC) and Ericsson.